\documentstyle[12pt,epsf]{article}
\newcommand{\be}{\begin{equation}}
\newcommand{\ee}{\end{equation}}
\newcommand{\bea}{\begin{eqnarray}}
\newcommand{\eea}{\end{eqnarray}}
\newcommand{\nn}{\nonumber}

\begin{document}

\def\gamh{\Gamma_H}
\def\esp #1{e^{\displaystyle{#1}}}
\def\de{\partial}
\def\eb{E_{\rm beam}}
\def\deb{\Delta E_{\rm beam}}
\def\sigm{\sigma_M}
\def\sigmmax{\sigma_M^{\rm max}}
\def\sigmmin{\sigma_M^{\rm min}}
\def\sige{\sigma_E}
\def\dsigm{\Delta\sigma_M}
\def\mh{M_H}
\def\lyear{L_{\rm year}}

\def\wstar{W^\star}
\def\zstar{Z^\star}
\def\ie{{\it i.e.}}
\def\etal{{\it et al.}}
\def\eg{{\it e.g.}}
\def\pzero{P^0}
\def\mt{m_t}
\def\mpzero{M_{\pzero}}
\def\mev{~{\rm MeV}}
\def\gev{~{\rm GeV}}
\def\gam{\gamma}
\def\lsim{\mathrel{\raise.3ex\hbox{$<$\kern-.75em\lower1ex\hbox{$\sim$}}}}
\def\gsim{\mathrel{\raise.3ex\hbox{$>$\kern-.75em\lower1ex\hbox{$\sim$}}}}
\def\ntc{N_{TC}}
\def\epem{e^+e^-}
\def\tauptaum{\tau^+\tau^-}
\def\lplm{\ell^+\ell^-}
\def\anti{\overline}
\def\mw{M_W}
\def\mz{M_Z}
\def\fbi{~{\rm fb}^{-1}}
\def\mupmum{\mu^+\mu^-}
\def\rts{\sqrt s}
\def\sigrts{\sigma_{\tiny\rts}^{}}
\def\sigrtssq{\sigma_{\tiny\rts}^2}
\def\sigrtsprime{\sigma_{E}}
\def\nsigrts{n_{\sigrts}}
\def\gampzero{\Gamma_{\pzero}}
\def\pzerop{P^{0\,\prime}}
\def\mpzerop{M_{\pzerop}}

\font\fortssbx=cmssbx10 scaled \magstep2
\hbox to \hsize{
%
%\special{psfile=uwlogo.ps
% hscale=8000 vscale=8000
% hoffset=-12 voffset=-2}
%\hskip.5in \raise.1in
%
$\vcenter{
\hbox{\fortssbx University of Florence}
%\hbox{\fortssbx University of Geneva}
}$
\hfill
$\vcenter{
\hbox{\bf DFF-343/8/99}
%\hbox{\bf UGVA-DPT-1999 05-xxx}
}$}

\medskip
\begin{center}

{\Large\bf\boldmath Effective Theory for Color-Flavor Locking in High
Density QCD\\}
\rm
\vskip1pc
{\Large R. Casalbuoni$^{a,b}$ and R. Gatto$^c$\\}
%\\ D. Dominici$^{a,b}$ S. De Curtis$^b$ and R. Gatto$^c$\\}
\vspace{5mm}
{\it{$^a$Dipartimento di Fisica, Universit\`a di Firenze, I-50125
Firenze, Italia
\\
$^b$I.N.F.N., Sezione di Firenze, I-50125 Firenze, Italia\\
$^c$D\'epart. de Physique Th\'eorique, Universit\'e de Gen\`eve,
CH-1211 Gen\`eve 4, Suisse}}
\end{center}
\bigskip
\begin{abstract}

\noindent
We describe the  low energy
excitations  of the  diquark condensates in
the color-flavor locked  phase of QCD with three massless flavors,
at high baryon densities,  in terms of a non-linear effective lagrangian.
Such a lagrangian
is formally seen to correspond
 to the lagrangian for the  hidden gauge symmetry description of the
effective low energy chiral
lagrangian,  the role of the hidden
symmetry being played by color.  In particular, this agrees with the
conjecture that the light
degrees of freedom of the two phases are in correspondence to each other
(complementarity).
The discussion includes consideration of the breaking of Lorentz invariance
in presence of
matter density.

\end{abstract}
\newpage
\section{Introduction}
We  introduce an effective lagrangian  describing the low energy
excitations of the  diquark condensates in
the color-flavor locked (CFL) phase of QCD with three massless flavors
at high chemical
potential. It is
expected that this phase is complementary to the low density chirality
breaking
phase by
identification of the light degrees of freedom in the two phases.
Within our effective description it is straightforward
to formally proof this
statement by showing that the effective color-gauged lagrangian in the CFL
phase
 is the non-linear lagrangian for the low density chirality breaking
phase
with explicit couplings to the octet of vector resonances. In fact we find
that the effective lagrangian in  the CFL phase corresponds
exactly to the so-called hidden gauge symmetry description of the effective
low energy chiral
lagrangian  with the role of the hidden
symmetry played by the (non hidden) gauge symmetry. The breaking of Lorentz
invariance
in presence of matter density does not hinder this conclusion.
The effective description we present here for color-flavor locking
constitutes a simplified
description of the full QCD dynamics. More quantitative details may only
come from a complete
dynamical treatment within QCD. Our approach exhibits however the main
features of the
color-locking phenomenon and it may be useful in view of its simplicity.

Understanding the behaviour of QCD in the domain of high densities and
small temperatures is
expected to be important for the study of neutron stars
and hopefully also for heavy ion collisions which are studied in
different high energy laboratories. Early studies \cite{ba} had suggested that
a new phenomenon, color superconductivity, would take place in those
conditions.

Particular attention has been given to the study of QCD with three massless
flavors
at high values of the baryonic chemical potential and zero temperature. At
very high
chemical potential one expects calculations based on one-gluon exchange to
be justified to derive the main traits of the superconductive dynamics. For
this case
the condensation pattern would imply the phenomenon of color-flavor locking
\cite{wilczek}.
In addition to the implied breaking of chiral symmetry in the locked phase,
gluons are
expected to become massive and an interesting correspondence takes place
between the
spectrum of the low lying states of the color-locked phase and
the hadronic spectrum in the chirally broken phase at low chemical
potential \cite{wilczek1}.
This feature (complementarity) will be a main interest in the present note.

Dynamical calculations of the superconductive gap have been carried out
by different authors \cite{different}. Beyond the ideal case of QCD with three
massless flavors, dynamical studies have also been attempted for two massless
flavors (u,d) and a massive s-quark, in the same region of high chemical
potential.  The construction we present here can easily
be modified to cover such a situation for the case of finite strange quark
mass
and sufficiently large chemical potential. Also the more complicated
situations
where both quark-antiquark and diquark condensates are present have been
examined. The extension of our effective lagrangian to such a possible
situation is still feasible although more cumbersome.

\section{The effective theory}

The CFL phase is characterized by the symmetry breaking pattern
\be
G=SU(3)_c\otimes SU(3)_L\otimes SU(3)_R \otimes U(1)\to
H=SU(3)_{c+L+R}
\ee
due to the dynamical formation of condensates of the type
\be
\langle\psi_{ai}^L\psi_{bj}^L\rangle=-\langle\psi_{ai}^R\psi_{bj}^R
\rangle=\kappa_1\delta_{ai}\delta_{bj}+\kappa_2\delta_{aj}\delta_{bi}
\ee
where $\psi_{ai}^{L(R)}$ are Weyl  spinors and a sum over spinor
indices is understood. The indices $a,b$ and $i,j$ refer to
$SU(3)_c$ and to $SU(3)_L$ (or $SU(3)_R$) respectively.

We are interested in building up a low energy effective description
of the CFL phase, that is, roughly, for
 momenta which are small with respect to the energy gap
characterizing this phase (the numerical estimates so far give values
between 10 and 100 $MeV$ for the gap energy).
The decay coupling constants of the Goldstones are
expected to be of the same order of magnitude of the gap.
In order
to describe this breaking pattern we need 17
Goldstone fields (for the moment being we will consider only
global symmetries and soon after come back to the gauging of color).

We introduce two types of coset fields transforming with respect to $G$
respectively as
a left-handed and a right-handed quark field. By denoting these
fields by $X$ and $Y$ we require
\be
X\to g_c X g_L^T,~~~~~Y\to g_c Y g_R^T
\ee
with
\be
g_c\in SU(3)_c,~~~g_L\in SU(3)_L,~~~g_R\in SU(3)_R
\ee
with  $X$ and $Y$ being $SU(3)$ matrices breaking respectively
$SU(3)_c\otimes SU(3)_L$ and $SU(3)_c\otimes SU(3)_R$.
This gives us 16
Goldstone fields. We then need an additional one to describe the
breaking of the baryon number. This is done in terms of a $U(1)$
factor transforming under $G$ as
\be
U\to g_{U(1)} U,~~~~g_{U(1)}\in U(1)
\ee
For the following it is convenient to define the following
currents
\be
J_X^\mu=X\de^\mu X^\dagger,~~~ J_Y^\mu=Y\de^\mu Y^\dagger,~~~
J_\phi=U\de^\mu U^\dagger
\ee
enjoying the following transformation properties with respect to
the global group $G$
\be
J_X^\mu\to g_c J_X^\mu g_c^\dagger,~~~J_Y^\mu\to g_c J_Y^\mu
g_c^\dagger,~~~J_\phi^\mu\to J_\phi^\mu
\ee
Notice that all these currents are anti-hermitian and
furthermore, since $J_X^\mu$ and $J_Y^\mu$ belong to the Lie
algebra of $SU(3)$, they are traceless. Then, barring  WZW terms
\cite{wess} (they are necessary for the completeness of the
theory but we will not discuss them here, except for a
comment at the end),
the most general Lorentz invariant lagrangian invariant under $G$, with at
most two
derivatives, is given by
\be
{\cal L}=-a Tr[J_X^\mu J_{X\mu}]-aTr[J_Y^\mu
J_{Y\mu}]-2cTr[J_X^\mu J_{Y\mu}^\dagger] -dJ_\phi^\mu J_{\phi\mu}
\label{lagrangian}
\ee

At this point we leave sub judice the question of Lorentz invariance
which does not hold in presence of matter density. We shall come back
soon after to this point.

Notice that the term proportional to $c$ is essential, otherwise
this lagrangian would be more symmetric than necessary, since it
would be invariant under $G\otimes SU(3)$. In fact, one could act
with independent transformations on the color indices of the
matrices $X$ and $Y$. It is just this term which allows to
identify the color groups acting upon these two fields. Such kind of term
is lacking
in ref.\cite{hong}. Also we
have required the theory to be invariant under parity, that is we
require symmetry under the exchange $X\leftrightarrow Y$.

For the
following it will be useful to write the previous lagrangian in
the form
\be
{\cal L}=-\frac{F^2}4 Tr[(J_X^\mu-J_Y^\mu)^2]-
\alpha\frac{F^2}4 Tr[(J_X^\mu+J_Y^\mu)^2] -\frac{f^2}2 (J_\phi^\mu)^2
\label{BESS}
\ee
where we have defined the parameters
\be
a-c=\frac{F^2} 2,~~~ a+c=\alpha\frac{F^2}2,~~~ d=\frac{f^2}2
\ee

Let us now define the Goldstone canonical fields. We start with
\be
X=\esp{i\tilde\Pi_X^aT_a},~~~
Y=\esp{i\tilde\Pi_Y^aT_a},~~~ U=\esp{i\tilde\phi},~~~
a=1,\cdots 8
\ee
and $T_a$  the generators of $SU(3)$ with the following
normalization (they are the Gell-Mann matrices up to a factor)
\be
Tr[T_aT_b]=\frac 1 2 \delta_{ab}
\ee
By defining
\bea
\Pi_X&=&\sqrt{\alpha}\,\frac{F}2(\tilde\Pi_X+\tilde\Pi_Y)\nn\\
\Pi_Y&=&\frac{F} 2(\tilde\Pi_X-\tilde\Pi_Y)\nn\\
\phi&=&f\tilde\phi
\label{rescaling}
\eea
we get at the lowest order in the Goldstone fields the properly
normalized kinetic term
\be
{\cal L}_{\rm kin}=\frac 12 \de_\mu\Pi_X^{a}\de^\mu\Pi_X^{a}+
\frac 12
\de_\mu\Pi_Y^{a}\de^\mu\Pi_Y^{a}+ \frac 1 2
\de_\mu\phi\de^\mu\phi
\ee
Therefore the expression of the original fields in terms of the
kinetic eigenstates is given by
\bea
X&=&\exp\left({\frac i F \left(\frac 1{\sqrt{\alpha}}\Pi_X^a+
\Pi_Y^a\right)T_a}\right)\nn\\
Y&=&\exp\left({\frac i F \left(\frac 1{\sqrt{\alpha}}\Pi_X^a-
\Pi_Y^a\right)T_a}\right)\nn\\ U&=&\exp\left({\frac i f\phi}\right)
\label{kin}
\eea
Let us now discuss the breaking of Lorentz invariance in
presence of a matter density. We have to
  treat differently  time
and spatial derivatives. In the lagrangian  (\ref{lagrangian}) we
should make substitutions of the type
\be
aTr[J^\mu_XJ^\dagger_{\mu X}]\to a_T Tr[J^0_XJ^\dagger_{0 X}]+
a_S Tr[J^i_XJ^\dagger_{i X}]
\ee
since we still have rotational invariance. The canonical term for
the Goldstone bosons is then
\bea
{\cal L}_{\rm kin}&=&\frac 1 2{\dot\Pi}_X^a {\dot\Pi}_X^a+
\frac 1 2{\dot\Pi}_Y^a {\dot\Pi}_Y^a+\frac 1 2\dot\phi\dot \phi\nn\\&-&
\frac 1 2 v_X^2{\vec\nabla}\Pi_X^a\cdot {\vec\nabla}\Pi_X^a-
\frac 1 2 v_Y^2{\vec\nabla}\Pi_Y^a\cdot {\vec\nabla}\Pi_Y^a-
\frac 1 2 v_\phi^2{\vec\nabla}\phi\cdot{\vec\nabla}\phi
\eea
where, with obvious notations
\be
v_X^2=\frac{\alpha_S F_S^2}{\alpha_T F_T^2},~~~v_Y^2=\frac{F_S^2}
{F_T^2},
~~~v_\phi^2=\frac{f_S^2}{f_T^2}
\label{velocity}
\ee
Therefore the three different types of Goldstone
bosons move with different velocities, but still satisfying a
linear dispersion relation $E=vp$. Of course, since we expect
here all the dimensioned quantities to be of the order of the
gap, we expect also the velocities to be of order 1. However,
since 8 of these Goldstone bosons must be eaten up by the gluons,
we will discuss again this point at the end of the next Section.

\section{The gauging of $SU(3)_c$}

The lagrangian of eq. (\ref{BESS}) must have a local $SU(3)_c$
invariance inherited by the color group of QCD. In order to make
this invariance explicit we need only to substitute the usual
derivatives with  covariant ones
\be
\de_\mu X\to D_\mu X= \de_\mu X-g_\mu X,~~~
\de_\mu Y\to D_\mu Y= \de_\mu Y-g_\mu Y,~~~ g_\mu\in {\rm Lie}~SU(3)_c
\ee
The corresponding currents are given by
\be
J_X^\mu=X\de^\mu X^\dagger+g^\mu,~~~ J_Y^\mu=Y\de^\mu
Y^\dagger+g^\mu
\ee
Therefore the lagrangian becomes
\bea
{\cal L}&=&-\frac{F^2} 4Tr[(X\de^\mu X^\dagger -Y\de^\mu
Y^\dagger)^2]-\alpha\frac{F^2} 4Tr[(X\de^\mu X^\dagger +Y\de^\mu
Y^\dagger+2g_\mu)^2]\nn\\ &-&\frac{f^2}2(J_\phi^\mu)^2+{\rm kinetic~
term~for}~ g^\mu
\label{lagr BESS}
\eea
Defining
\be
g_\mu=ig_s\frac{T_a}2 g_\mu^a
\ee
where $g_s$  is the QCD coupling constant, we see that the gluon
field acquires a mass (more discussion later). In the limit where the
Goldstones have energies much smaller than the gluon mass, we can
neglect the kinetic term for the gluon. Then the lagrangian
(\ref{lagr BESS}) is nothing but the hidden gauge symmetry
version of the chiral lagrangian for QCD \cite{BESS} (except for
the contribution of the field $\phi$). In fact, in this limit,
the gluon field becomes an auxiliary field which can be
eliminated through its equation of motion
\be
g_\mu=-\frac 1 2(X\de_\mu X^\dagger+Y\de_\mu Y^\dagger)
\ee
obtaining
\be
{\cal L}=-\frac{F^2}4Tr[(X\de_\mu X^\dagger-Y\de_\mu
Y^\dagger)^2]-\frac{f^2} 2(J_\phi^\mu)^2=\frac{F^2}4 Tr[\de_\mu \Sigma\de^\mu
\Sigma^\dagger] -\frac{f^2} 2(J_\phi^\mu)^2
\label{QCD}
\ee
where
\be
\Sigma=Y^\dagger X
\ee
transforms under the group $SU(3)_c\otimes SU(3)_L\otimes
SU(3)_R$ as
\be
\Sigma\to g_R^* \Sigma g_L^T
\ee
The field $\Sigma^T$ transforms exactly as the standard chiral field
of the non linear-realization of QCD (in the chiral broken
phase).

The preceding discussion applies to any situation for which the mass of the
gluon
is at least larger of the typical gap parameter. For a situation in which
the mass
of the gluon is much smaller than the gap parameter the degrees of freedom
of the massive gluons have to kept in the effective lagrangian. For such a
situation
we can make  use of the gauge  freedom to
choose a gauge such that $X=Y^\dagger$. This implies
\be
\tilde\Pi_X=-\tilde\Pi_Y
\ee
or
\be
\Pi_X=0,~~~\Pi_Y=F\tilde\Pi_X
\ee
This gauge is the unitary gauge. In fact the bilinear term in the
Goldstones and in the gluon field in eq. (\ref{lagr BESS}) is
proportional to
\be
g^\mu\de_\mu(\tilde\Pi_X+\tilde\Pi_Y)
\ee
and therefore cancels out. It follows that the gluon mass is
given by
\be
m_g^2=\alpha g_s^2\frac {F^2}4
\ee
Notice also that  the $X\leftrightarrow Y$ symmetry, in this
gauge, implies $\Pi_Y\leftrightarrow -\Pi_Y$.

The identification
with the low-energy effective lagrangian of QCD could not seem to
be really correct, since in this phase we are breaking Lorentz
invariance. However in the low-energy limit, where only the
fields $\Sigma$ and $\phi$ are present,  this effect amounts to
take as lagrangian
\be
{\cal L}=\frac{F_T^2}4 \left(Tr[\dot \Sigma\dot \Sigma^\dagger]- v_Y^2
Tr[\de^i \Sigma\de^i \Sigma^\dagger]\right)
-\frac{f_T^2}2\left((J_\phi^0)^2-v_\phi^2(J_\phi^i)^2\right)
\ee
We see that by a rescaling of the coordinates $x^i\to v_Y x_i$
(that is by using a coordinate system with the $v_Y=1$), the
first term becomes exactly the chiral lagrangian for QCD.

 The question  remains of the Goldstone $\phi$
associated to the breaking of the baryon number. According to
ref. \cite{wilczek1} this could be interpreted as a dibaryon
state, the $H$, a spin 0 $SU(3)$ singlet of composition
$(udsuds)$. In fact, it had been pointed out by R. Jaffe \cite{jaffe},
 that this (deus ex machina) state is particularly light.

\section{Gauging $U(1)_{\rm em}$}

The gauging of  $U(1)_{\rm em}$ (formal extension to the full electroweak
group
is straightforward) is made through covariant
derivatives
\be
D_\mu X=\de_\mu X-g_\mu X -X Q A_\mu,~~~ D_\mu Y=\de_\mu Y-g_\mu
Y -Y Q A_\mu
\ee
where $Q={\rm diag}(+2/3,-1/3,-1/3,)$. We see immediately that
the combination of $Q$ and of the generator of $SU(3)_c$
proportional to   $Q_{SU(3)_c}={\rm diag}(-2/3,+1/3,+1/3)=-Q$,
leaves invariant the ground state. In fact,
\be
Q_{SU(3)_c}\langle X\rangle+\langle X\rangle Q
\to ({Q_{SU(3)_c}})_{ab}\delta_{bi}+\delta_{ai}Q_{ij}=0
\ee
It follows that one combination of $g^Q$ (the gluon field
associated to $Q_{SU(3)_c}$) and $A_\mu$ is massive, whereas the
orthogonal combination remains massless. Since the component of the
gluon field involved in the mixing, with the appropriate
normalization, is given by
\be
g_\mu\to \frac i 2 \sqrt{\frac 32} g_s g_{\mu }^Q Q
\ee
and $A_\mu$ is defined as
\be
A_\mu\to ie A_\mu
\ee
one  easily finds that the massless combination is
\be
A'=A\cos\psi+g_Q\sin\psi
\ee
with
\be
\tan\psi=\sqrt{\frac 83}\frac e{g_s}
\ee
As a last comment, let us consider at this order in the
derivatives the WZW term associated to the $U(1)$ anomaly. It
will be proportional to
\be
Tr(\tilde
Q^2\Pi_Y)\epsilon_{\mu\nu\rho\sigma}F^{\mu\nu}F^{\rho\sigma}
\ee
The calculation of the trace can be done by evaluating the
contribution of the quarks contributing in the loop, by using the
definition of the electric charge in this basis.
In an equivalent way we can use the
previous expression, being careful that the action of the electric
charge in this phase is really of the type
\be
\tilde Q=1\otimes Q-Q\otimes 1
\ee
where $Q={\rm diag}(+2/3,-1/3,-13)$ and the tensor product refers
to the color and flavor indices respectively. Then, by taking the
trace with the $\Pi_Y$ field (acting on the flavor space since it
comes from $\Sigma$), we get easily
\be
Tr[\tilde Q^2 \Pi_Y]=3Tr[Q^2\Pi_Y]
\ee
Therefore, also in this phase we get the anomalous decay of the
two mesons lying in the center of $SU(3)$, since the previous
trace gets contribution both from $T_3$ and from $T_8$, as it can
be seen by using the expression of $Q$ in terms of the $SU(3)$
generators
\be
Q=\frac{\sqrt{3}} 2\left(T_3+\frac 1 {\sqrt{3}} T_8\right)
\ee
This calculation gets much simpler by using a basis of flavor
where the indices $i=1,2,3$ correspond to $s,d,u$ respectively,
since in this basis $Q$ is proportional to $T_8$.

\section{Conclusions}
We have shown that QCD for three massless flavors in  the color-flavor locked
 (CFL) phase at high densities
is described at low energies by an
effective lagrangian which is the same as the low energy effective
lagrangian of the chiral phase. The gluons become massive in the CFL phase
and the effective lagrangian of such a phase, when
reinterpreted in the chiral phase, describes an octect of
vectors coupled in a chiral invariant way, allowing for the
interpretation of the massive gluons as  corresponding to the
ordinary vector bosons. The lack of Lorentz invariance in the dense phase does
not affect the argument.

\begin{center}
{\bf Acknowledgements}
\end{center}
\medskip
R.C.  would like to thank the Theory Division of CERN for the
kind hospitality offered to him during the final stage of this
paper.

\end{document}